# Time-resolved mid-infrared dual-comb spectroscopy

**Muhammad A. Abbas, Qing Pan, Julien Mandon, Simona M. Cristescu, Frans J. M. Harren & Amir Khodabakhsh***

Trace Gas Research Group, Department of Molecular and Laser Physics, Institute for Molecules and Materials, Radboud University, 6525 AJ Nijmegen, The Netherlands

## Abstract

Dual-comb spectroscopy can provide broad spectral bandwidth and high spectral resolution in a short acquisition time, enabling time-resolved measurements. Specifically, spectroscopy in the mid-infrared wavelength range is of particular interest, since most of the molecules have their strongest rotational-vibrational transitions in this "fingerprint" region. Here we report time-resolved mid-infrared dual-comb spectroscopy for the first time, covering ~300 nm bandwidth around 3.3 μm with 6 GHz spectral resolution and 20 μs temporal resolution. As a demonstration, we study a $CH_4$/He gas mixture in an electric discharge, while the discharge is modulated between dark and glow regimes. We simultaneously monitor the production of $C_2H_6$ and the vibrational excitation of $CH_4$ molecules, observing the dynamics of both processes. This approach to broadband, high-resolution, and time-resolved mid-infrared spectroscopy provides a new tool for monitoring the kinetics of fast chemical reactions, with potential applications in various fields such as physical chemistry and plasma/combustion analysis.

Time-domain monitoring of fast chemical reactions is of particular interest in several fundamental and applied scientific fields, including physical chemistry, plasma/combustion analysis, biology, and environmental studies. Broadband, time-resolved absorption spectroscopy can provide the possibility to simultaneously monitor time-dependent parameters of the chemical reactions, such as concentrations of intermediate/final chemical products, transient free radicals and ions, as well as branching ratios, reaction rate coefficients, temperature and number densities of molecular excited-states. Generally, the main challenge is to obtain a broadband spectrum with high spectral resolution and high detection sensitivity in a short measurement time. Continuous-wave (cw) laser absorption spectroscopy can provide time-resolved measurements for a single chemical species with a high detection sensitivity. However, for a broad spectral coverage the laser source needs to be scanned over the spectral range, inevitably reducing the measurement speed. Alternatively, one can use broadband time-resolved absorption spectroscopy techniques, which are traditionally based on incoherent light sources. They can provide an ultra-broadband time-resolved spectrum, but they need a long averaging time to achieve a high signal-to-noise ratio (SNR) and detection sensitivity. Two widely used methods are step-scan mechanical Fourier transform spectroscopy (FTS) [1-3] and dispersion-based detection [4-7]. The former exhibits very long measurement times due to the step-scanning, while the latter yields shorter measurement times, but usually has a coarse spectral resolution.

In contrast to these traditional broadband methods, optical frequency comb spectroscopy (OFCS) simultaneously provides a broad spectral coverage and a high spectral resolution. It can also yield a high SNR within a short measurement time, due to the coherency and high spectral brightness of optical frequency comb sources. Specifically, OFCS in the mid-infrared (mid-IR) wavelength range (2-20 µm) has been of particular interest, since almost all molecules have their fundamental rotational-vibrational transitions in this region with distinct absorption patterns (i.e. fingerprints). Various OFCS techniques have been utilized in the mid-IR wavelength region; e.g. combining an optical frequency comb with a mechanical FTS [8], dual-comb spectroscopy (DCS) [9-11], and dispersion-based methods [12-15]. A comprehensive review of these spectroscopic methods can be found elsewhere [16].

Time-domain/time-resolved spectroscopy using optical frequency combs has emerged strongly in the last decade. In a first demonstration, DCS was used for measuring molecular free induction decay in the near-infrared (near-IR) wavelength range using two Er:fiber mode-locked lasers [17]. A few other works have been reported in near-IR region using Ti:sapphire mode-locked lasers including multidimensional dual-comb spectroscopy (M-DCS$^2$) able to differentiate and assign the Doppler-broadened features of two naturally occurring isotopes of Rb [18], dual frequency comb-based transient absorption (DFC-TA) spectroscopy for measurement of the relaxation processes of dye molecules in solution from femtosecond to nanosecond timescales [19], and DCS for the study of laser-induced plasma from a solid sample, simultaneously measuring trace amounts of Rb and K in a laser ablation [20]. In the visible range (~530 nm), cavity-enhanced transient absorption spectroscopy (CE-TAS) has been demonstrated for study of the ultrafast dynamics of $I_2$ in a molecular beam [21], and more recently, time-resolved dual-comb spectroscopy has been reported for measurement of number density and temperature in a laser-induced plasma by monitoring three excited-state transitions of Fe [22]. In the mid-IR region, the demonstration has been limited to cavity-enhanced time-resolved frequency comb spectroscopy (TRFCS) for monitoring of transient free radicals and kinetics of the OD + CO → DOCO reaction by 2D cross-dispersion of the spectrum on a liquid $N_2$ cooled camera using a virtually imaged phase array (VIPA) etalon in combination with a conventional grating [23-25].

Here, we report time-resolved dual-comb spectroscopy (TRDCS) in the mid-IR wavelength region, for the first time to the best of our knowledge, and demonstrate its application for monitoring fast chemical dynamics. For this, we study the vibrational excitation/de-excitation of $CH_4$ in an electrical discharge and the concentration of the reaction product $C_2H_6$, at millisecond and microsecond time scales, while the discharge is modulated between dark and glow regimes.

## Results

**Dual-comb mid-IR spectrometer.** The dual-comb spectrometer (Fig. 1) is based on a femtosecond singly-resonant optical parametric oscillator (OPO) containing two nonlinear crystals in a single cavity. The crystals are synchronously pumped by two Yb:fiber mode-locked lasers (counter propagating and cross polarized) with a stabilized repetition rate ($f_{rep}$) difference of $\Delta f_{rep} \approx 250$ Hz and free running carrier-envelope offset frequencies ($f_{ceo}$) [10,26]. The two mid-IR idler beams (~3.3 μm) are combined on a beam splitter producing two pairs of beams. One pair is transmitted through a ~50 cm-long discharge tube (diameter 3 mm) containing $CH_4$ diluted in He, and focused onto a fast photodetector. The discharge tube has a continuous gas flow (4 normal liter/hour, 4 Nl/h) and is connected to a gas handling system. The second comb pair is sent through a reference absorption cell (filled with $CH_4$ at low pressure) and dispersed by a diffraction grating. A part of the spectrum is focused on a second photodetector to monitor a single well-defined absorption line of the reference sample. The time-domain interferograms in the output of the two photodetectors are digitized, a Blackman apodization function is applied to the both (sample and reference) interferograms, and followed by a Fourier transformation to yield the corresponding absorption spectra. The apodization function is used to minimize the ringing effect around the (narrow) absorption lines [27], which also limits the effective measurement time-window of each interferogram to ~120 μs around the central burst. Therefore, each individual interferogram is measured in 120 µs with a repetition period of $1/\Delta f_{rep}$ ($\approx 4$ ms). The absorption line in the spectrum of the reference gas is used to correct for the frequency jitter in the sample spectrum, which is mainly due to the free running $f_{ceo}$ values; it also provides an absolute optical frequency calibration. The free-running $f_{ceo}$ of the two combs makes the experimental setup much less complex compared to the state-of-the-art, fully stabilized, mid-IR dual-comb spectrometers [28,29] . The explanation of the data acquisition and signal processing of the optical referencing method can be found elsewhere [27] and a more detailed description of the experimental setup is presented in the Methods section.

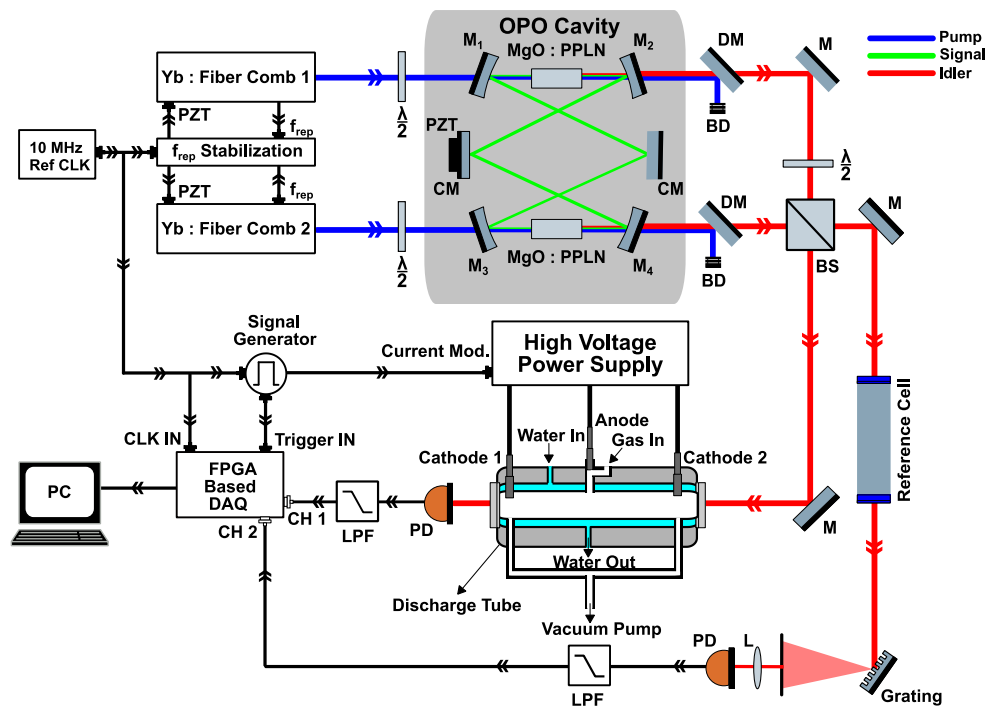

Fig. 1. Experimental setup. Two femtosecond Yb:fiber lasers with stabilized and slightly different $f_{rep}$ (and free running $f_{ceo}$), synchronously pump two MgO:PPLN crystals in a single OPO cavity, providing two mid-IR idler beams. The two combined idler beams are sent through the sample (discharge) cell and a reference gas cell. The latter yields the dual-comb spectrum of a single well-defined

absorption line, which is used to correct for the free running $f_{ceo}$, as well as absolute frequency calibration of the sample spectrum. *M* flat mirror, $M_{1-4}$ concave dielectric mirror, *CM* flat chirped dielectric mirror, *DM* dichroic mirror, *BD* beam dump, *BS* beam splitter, *λ/2* half-waveplate, *L* lens, *PD* photodetector, *LPF* low pass filter.

To demonstrate the performance of the spectrometer, we filled the discharge cell with 50% $CH_4$ diluted in He, at 25 mbar total pressure and a total flow rate of 4 Nl/h, and measured the transmission spectrum without the discharge. Figure 2a shows the normalized transmission spectrum of the fundamental $CH_4$ transition from the ground state to the $\nu_3$ vibrational state (in black, 500 averages, ~2 s measurement time). The spectrum is normalized to an averaged background spectrum, which was recorded when the sample absorption cell was filled with pure He at 25 mbar. We compare this experimental absorption spectrum to the theoretical model spectrum of $CH_4$ (in blue) developed based on the corresponding parameters from the HITRAN database [30]. We used a Voigt profile and convoluted the model spectrum with a Blackman instrument line-shape function, corresponding to the applied apodization function. Note that the simulated spectrum is inverted and the two measured and simulated spectra are offset for clarity. We fit the developed model spectrum to the measured spectrum (with $CH_4$ concentration as the fitting parameter) and also take into account the remaining baseline and etalon fringes by including a sum of a low order polynomial and few low frequency sinewave functions in the fit. The retrieved $CH_4$ concentration is 49.7(9)%, where the error is the standard deviation of 10 consecutive measurements. The spectral resolution of the measured spectrum is ~6 GHz (~0.2 $cm^{-1}$), and the precision of the frequency calibration is ~120 MHz (~0.004 $cm^{-1}$). The residual of the fit is shown in Fig. 2b, indicating the good agreement between the measurement and the model.

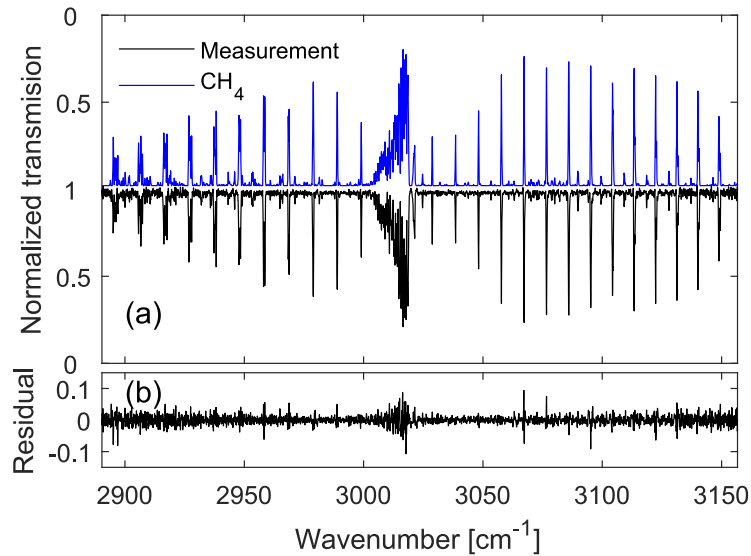

Fig. 2. Measured spectrum without discharge. (a) Normalized transmission spectrum of 50% $CH_4$ diluted in He at 25 mbar total pressure and room temperature (black, 500 averages), along with a fit model spectrum of $CH_4$ (in blue, inverted and offset for clarity) using HITRAN database parameters. (b) Residual of the fit.

**Methane spectrum in a static discharge.** We measured the normalized transmission spectrum of the aforementioned gas sample (50% $CH_4$ in He, 25 mbar) in a static discharge. We applied a DC voltage of 10 kV and a stabilized current of 10 mA (current density 1.4 mA/mm$^2$) to the discharge tube and kept a constant sample gas flow of 4 Nl/h through the cell. Figure 3a shows the measured normalized spectrum of the sample with the discharge (in red, 500 averages, ~2 s measurement time) compared to the measured normalized spectrum of the sample without the discharge (in black, 500 averages, ~2 s measurement time). Note that, neither a baseline correction nor an etalon-fringes removal process is applied to these spectra. In the discharge the $CH_4$ absorption is reduced, which indicated a lower population in the vibrational ground state, due to vibrational excitation, ionization and molecular dissociations. This is accompanied by the appearance of two groups of additional absorption lines in the spectrum with the discharge, enlarged in Fig. 3b and Fig. 3c. The additional absorption lines in Fig. 3b are due to the produced $C_2H_6$ in the discharge. Since the $C_2H_6$ model in the HITRAN database is incomplete, we measured the room temperature absorption spectrum of 10% $C_2H_6$ diluted in He (25 mbar total pressure) in order to obtain a proper reference spectrum for $C_2H_6$. We fit the retrieved reference spectrum to the recorded discharge spectrum, with the concentration of $C_2H_6$ as the fitting parameter. This fitted spectrum is shown in Fig. 3b (in blue, inverted for clarity), and the retrieved $C_2H_6$ concentration from the fit is 6.52(11)%. Note that, we excluded few $C_2H_6$ absorption lines from the broadband fitting routine, since they showed deviation from the corresponding absorption lines in the reference spectrum, most probably due to differences in number densities at these levels caused by the discharge. The second group of additional absorption lines, indicated by "*" in Fig. 3c, are due to vibrational hot band transitions in $CH_4$, since $CH_4$ molecules are excited in the discharge. Many more (weaker) hot band absorption lines appear in all three P-, Q-, and R- branches, but are not highlighted in the figure. The first vibrational excited state of $CH_4$ is $\nu_4$ (energy level ~1310 $cm^{-1}$) followed by $\nu_2$ (energy level ~1533 $cm^{-1}$). The majority of the detected hot band absorption lines undergo a vibrational $\nu_3+\nu_4$ $\leftarrow \nu_4$ transition and a smaller portion originate from the $\nu_3+\nu_2 \leftarrow \nu_2$ transition. All hot band ro-vibrational transitions can be assigned using the HITRAN database.

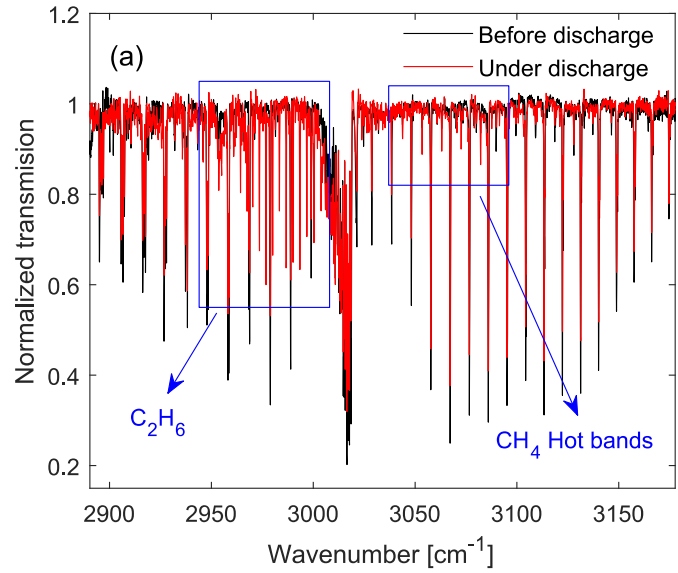

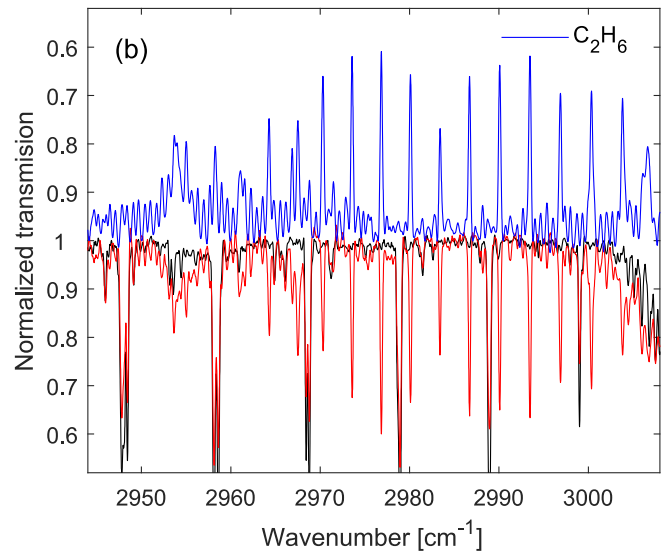

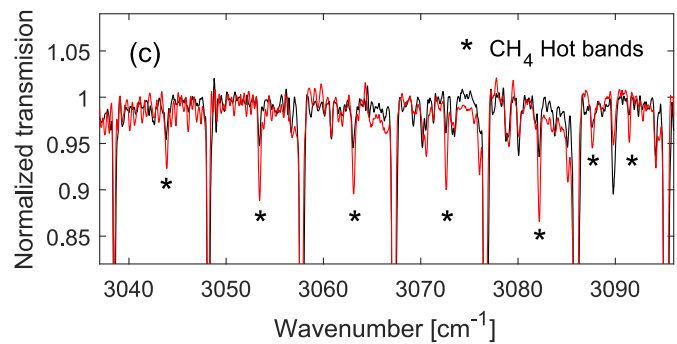

Fig. 3. Methane spectrum in the discharge. (a) Normalized spectrum with the discharge (in red, 500 averages) compared to the normalized spectrum without the discharge (in black, 500 averages). (b) An enlargement to the absorption lines of $C_2H_6$ produced in the discharge compared to a fitted room temperature spectrum of $C_2H_6$ (in blue, inverted). (c) An enlargement to a number of stronger rotational lines (indicated by '*') within the hot band transitions of $CH_4$.

**Time-resolved measurement at milliseconds time scale.** To perform time-resolved measurements, we modulated the current of the discharge, with a square-wave function in an "off" (dark) and "on" (glow) regime (20% duty cycle for "on" time). The current modulation was synchronized with the repetition rate difference ($\Delta f_{rep} \approx 250$ Hz) of the dual-comb spectrometer, and the modulation frequency, $f_{mod}$, was chosen to be equal to $\Delta f_{rep}$ divided by 100, i.e. $f_{mod} = \Delta f_{rep}/100$ ($\approx 2.5$ Hz). Therefore, each period of the discharge modulation was recorded by a set of 100 consecutive interferograms, each at its own time-bin. This allowed averaging of the spectra (after Fourier transform of the interferograms) for each corresponding time-bin of different discharge cycles to achieve high SNR averaged spectra. Therefore, we monitored each period of the discharge modulation with a time resolution equal to the inverse of $\Delta f_{rep}$, i.e. $T_{res} = 1/\Delta f_{rep}$ ($\approx 4$ ms). We obtained the normalized spectra of the gas sample (50% $CH_4$ in He, 25 mbar, flow rate 4 Nl/h) over the entire period of the modulation (400 ms) with a time-resolution of 4 ms. Figure 4a shows the retrieved concentrations of the generated $C_2H_6$, and Fig. 4b demonstrates the absorbance of R(7) line at ~3082.2 $cm^{-1}$ corresponding to the vibrational $\nu_3+\nu_4 \leftarrow \nu_4$ hot band transition of $CH_4$, as an indicator of the number of excited $CH_4$ molecules. The discharge was turned on and off at t = 0 ms and t = 80 ms, respectively. Each data point is averaged for 250 times (recording time of 1 s) and the total measurement time for the entire data set is ~100 s. Note that we began to record the data after the modulation was applied for a few minutes, in order to avoid any possible transient conditions from a static to a dynamic regime.

As shown in Fig. 4a, after turning the discharge on at t = 0 ms, the concentration of $C_2H_6$ increases from 0.71% to 2.4% in ~20 ms and reaches to a pseudo-plateau region. The absorbance of the hot band line abruptly appears at t = 0 ms, as shown in Fig. 4b, not resolvable with the current (4 ms) time resolution. After this abrupt appearance, the absorbance demonstrates a decrease to half of its initial value in ~20 ms and reaches to a pseudo-plateau region. The comparable time scale in the increase of the $C_2H_6$ concentration and decrease of the hot band absorbance suggests that these two processes are likely to be correlated. After the discharge is turned off at t = 80 ms, the $C_2H_6$ concentration increases instantaneously from 2.5% to 6.7%, while the hot band transition disappears. None of these two processes can be resolved with 4 ms time resolution. After the sudden increase, the $C_2H_6$ concentration decreases linearly, within the gas refresh time of the long and narrow discharge cell, before it reaches a pseudo-plateau region slowly decreasing from 1.0% to 0.71% by the end of the cycle. The latter demonstrates the remaining $C_2H_6$ molecules in the discharge cell most probably due to purging inhomogeneity.

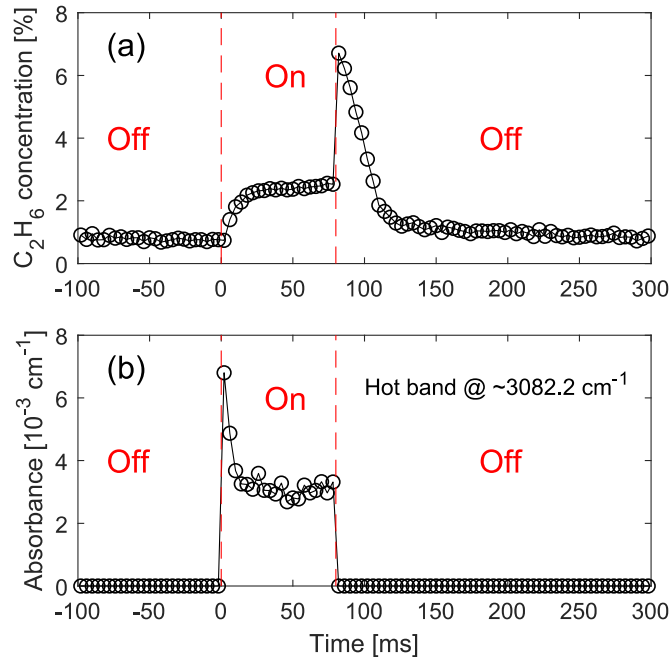

Fig. 4. Time-resolved measurement at milliseconds time scale. (a) Concentrations of generated $C_2H_6$ and (b) absorbance of the R(7) rotational line at ~3082.2 cm$^{-1}$ in the vibrational $\nu_3+\nu_4 \leftarrow \nu_4$ hot band of $CH_4$, with 4 ms time resolution, recorded while the discharge is switched between dark and glow regimes. Different states of the discharge are separated by red dashed lines and are indicated by red "On" and "Off" labels.

**Time-resolved measurement at microseconds time scale.** To perform time-resolved measurement at microseconds time scale, the square-wave current modulation was synchronized with $\Delta f_{rep}$, with an equal frequency $f_{mod} = \Delta f_{rep}$ ($\approx 250$ Hz), while the current modulation could be deliberately time-delayed (phase-shifted) with respect to $\Delta f_{rep}$. For each particular time-delay the consecutive interferograms were recorded and directly averaged after Fourier transform to yield a high SNR spectrum. To cover the period of the current modulation, the time-delay was step-scanned and an averaged spectrum was measured for each step. In this configuration the time resolution is equal to the time-delay steps (e.g. 20 μs); however, for rapid changes happening faster than the measurement time of each individual interferogram (120 μs), the measurement results would be smoothened by a moving average. We obtained the normalized spectra of 50% $CH_4$ in He (25 mbar, flow rate 4 Nl/h) over the entire discharge modulation period of 4 ms (square-wave function, ~10% duty cycle for "on" time). The step size was 20 μs near the switching events and 50 μs far from the switching events. Figure 5a shows the generated $C_2H_6$ concentrations and Fig. 5b demonstrates the absorbance of the same $CH_4$ hot band line of R(7) at ~3082.2 $cm^{-1}$, that was previously monitored in the millisecond time scale. Data are shown for a time span of 1 ms around the 400 μs period that the discharge was on. Each data point is averaged for 250 times (recording time of 1s) and the total measurement time for the shown data set is ~42 s, excluding the standby time for varying the time-delay.

In contrast to the results obtained in the milliseconds time scale, no discontinuity of the monitored parameters is observed in the microseconds time scale measurements. Note that after applying the current modulation to the discharge, the system operated for a few minutes before recording the data, to avoid any possible transient conditions from a static to a dynamic regime. Due to the high frequency of the discharge modulation the gas flow is insufficient to purge the generated $C_2H_6$ on each modulation cycle, which leads to an accumulation of $C_2H_6$ up to a concentration of 7.7% just before the discharge is turned on, as shown in Fig. 5a. When the discharge is turned on at t = 0 μs, the $C_2H_6$ concentration initially decreases rapidly to 3.0% in ~100 μs, after which it increases slightly to 3.2% in the next 300 μs. This slight increase reflects early-time dynamics, observed at the milliseconds time scale measurements right after turning the discharge on. In Fig. 5b, the absorbance of the monitored hot band line demonstrates a rapid increase to a comparable amplitude that has been observed with the milliseconds time scale measurements. The rise time is ~60 μs, which is faster than the observed fall time of the $C_2H_6$ concentration. After the rapid formation of the hot band line, the absorbance amplitude is slowly decreased from ~$7.5 \times 10^{-3}$ cm$^{-1}$ to ~$6.5 \times 10^{-3}$ cm$^{-1}$, right before the discharge is turned off. This is in agreement with the observed decay dynamics in the milliseconds time scale measurement, following the initial abrupt increase. After the discharge is turned off, the $C_2H_6$ concentration increases from 3.2% to 7.6 % in ~500 μs. Meanwhile, the absorbance of the monitored hot band line decreases to zero in ~200 μs. The dynamics of the vibrational excited $CH_4$ and the generated $C_2H_6$ are opposing each other, with the former slightly faster than the latter. The dynamics of $C_2H_6$ implies that when the discharge is turned on, it dissociates the $C_2H_6$ into free radicals, such as $C_2H_5 \cdot$ and $CH_3 \cdot$, while after the discharge is turned off these radicals recombine, and amongst others, form $C_2H_6$. This interpretation also explains the abrupt increase of $C_2H_6$ concentration after the discharge was turned off in the milliseconds time scale measurements (Fig 4b at t = 80 ms).

At these pressures in the discharge, we can also observe that the dissociation (formation) of $C_2H_6$, to (by recombination of) free radicals is slower than vibrational excitation (de-excitation) of $CH_4$ molecules. The vibrational de-excitation of methane is mainly due to the collisions. We measured the collisional de-excitation decay time of different rotational lines in the hot bands, by fitting exponential functions to the absorbance of the hot band lines (after the discharge is turned off).

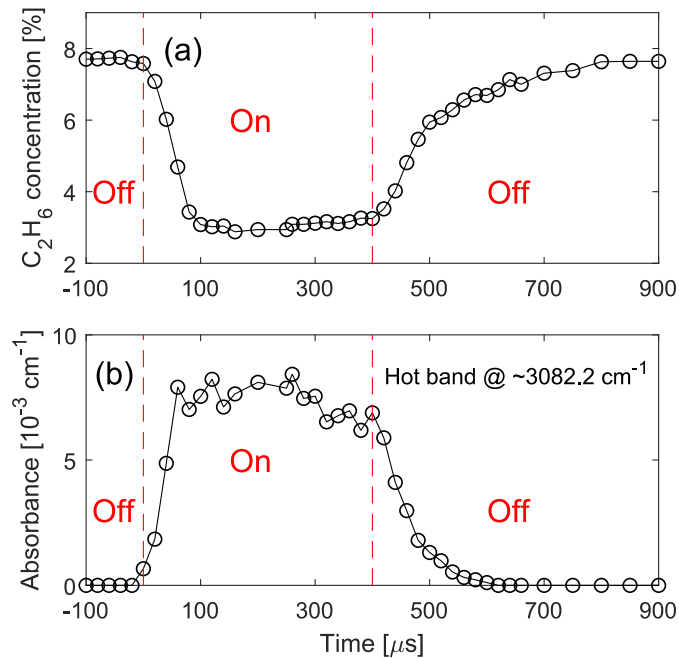

Fig. 5. Time-resolved measurement at microseconds time scale. (a) Concentrations of generated $C_2H_6$ and (b) absorbance of the R(7) rotational line at ~3082.2 cm$^{-1}$ in the vibrational $\nu_3+\nu_4 \leftarrow \nu_4$ hot band of $CH_4$, with 20 μs time resolution, recorded while the discharge is switched between dark and glow regime. Different states of the discharge are separated by red dashed lines and are indicated by red "On" and "Off" labels.

Figure 6 shows the corresponding measurements and fits for absorbance of three rotational hot band lines, R(2) at ~3043.9 cm$^{-1}$ and R(7) at ~3082.2 cm$^{-1}$ corresponding to the vibrational $\nu_3+\nu_4 \leftarrow \nu_4$ hot band transition, and R(8) at ~3091.4 cm$^{-1}$ corresponding to the vibrational $\nu_3+\nu_2 \leftarrow \nu_2$ hot band transition. The hot band ro-vibrational transitions were assigned using the HITRAN database. The collisional de-excitation decay times retrieved from the fits are 57(5) μs, 52(3) μs, and 54(2) μs, respectively. Although the decay times are shorter than the measurement time of each individual interferogram (120 μs), they are not affected by the moving average, since the shape and decay rate of an exponential function will not be affected by integration. This is also evident by the good agreement between the measurements and the fit exponential functions.

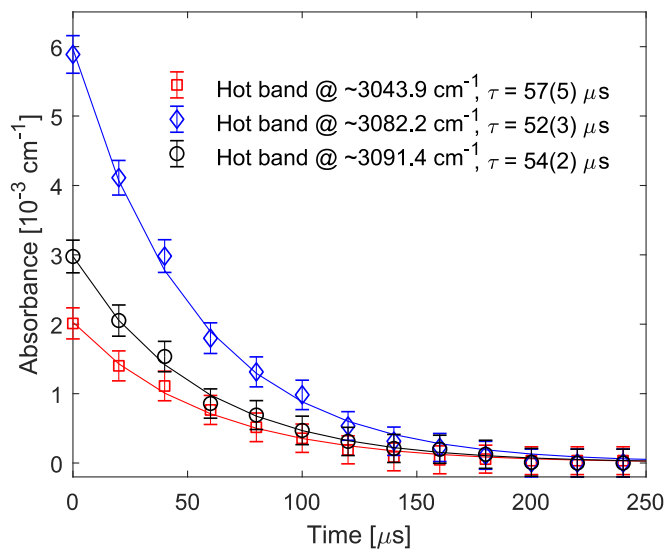

Fig. 6. Collisional de-excitation of three hot band lines. The rotational lines are R(2) at ~3043.9 cm$^{-1}$ (red squares) and R(7) at ~3082.2 cm$^{-1}$ (blue diamonds) in the vibrational transition of $\nu_3+\nu_4 \leftarrow \nu_4$, as well as R(8) at ~3091.4 cm$^{-1}$ (black circles) in the vibrational transition of $\nu_3+\nu_2 \leftarrow \nu_2$ of CH$_4$. The time-resolved absorbance is demonstrated along with an exponential fit to each data set. The retrieved decay times form the fits are shown in the figure legend.

## Conclusion

We demonstrated, for the first time, the capabilities of time-resolved dual-comb spectroscopy (TRDCS) in the mid-IR wavelength range. For this, we studied the vibrational excitation and de-excitation of CH$_4$ in a modulated electrical discharge and its reaction product C$_2$H$_6$ at milliseconds and microseconds time scales. The total acquisition time for each measurement was in the order of tens of seconds. The spectrometer covered a wavelength range from ~3.1 μm to ~3.4 μm with a spectral resolution of 6 GHz and a single shot acquisition time of ~120 μs. By tuning the OPO light source, it is possible to monitor the other wavelength ranges from 2.7 μm to 4.7 μm. In this initial demonstration of TRDCS, we observed the products at %-levels due to the relative short interaction path length. The detection sensitivity of the spectrometer can be enhanced by using a multipass arrangement or an enhancement resonant cavity. Higher detection sensitivity, in combination with the broad spectral tunability of the OPO, provides the possibility to monitor less abundant interesting species in the discharge process in different wavelength ranges, e.g. free radicals, transient species and ions. In the dynamics, a wealth of information becomes available, which we only scratched the surface in this first demonstration. A comprehensive time-resolved study of a discharge process for the kinetics of reactions and branching next to temperature information could be feasible. The system can also be used for studying any fast chemical reaction that can be periodically triggered with an acceptable level of reproducibility. Recently, TRDCS in the near-IR wavelength range has been utilized to monitor a fast, single shot reaction with millisecond time scale resolution [31]. The results are promising, although they are limited to high pressure spectroscopy for the time being. Finally, it should be noted that DCS has an inherent trade-off between the measurement time and the spectral resolution, which provides flexibility to choose a proper combination for each particular application. The minimum measurement time can be deliberately selected to be the shortest time with which the spectral resolution is still sufficient for detecting the desired species and/or resolving the target spectral features.

## Methods

**Optical setup.** The two mid-IR frequency combs are generated from a singly-resonant optical parametric oscillator (OPO) based on two MgO:PPLN crystals (Covesion Ltd., UK) positioned in a common OPO cavity. Each crystal is synchronously pumped by a Yb:fiber mode-locked laser (Menlo Systems, Germany) emitting around 1040 nm. The pump beams are counter propagating in the ~3.3 m-long OPO cavity and are perpendicularly polarized. The repetition rates ($f_{rep}$) of the mode-locked lasers are ~90 MHz and stabilized to a common reference clock but are slightly different ($\Delta f_{rep} \approx 250$ Hz), yielding the repetition rate difference in the generated idlers combs. The carrier-envelop offset frequency ($f_{ceo}$) of the two pump mode-locked lasers as well as the OPO cavity length are not stabilized. Each idler beam can provide up to 200 mW of average power, covering a wavelength range of around 350 nm, and is tunable from 2.7 to 4.7 μm using different poling periods in the nonlinear crystals. The two idler beams (after polarization adjustment) are combined by a beam splitter to produce two pair of beams on reflection and transmission. One pair is transmitted through the discharge tube, filled with $CH_4$ diluted in He at a total pressure of 25 mbar. The transmitted beams are focused on a fast (50 MHz) thermoelectrically cooled HgCdTe photodetector (PVI-4TE, Vigo System, Poland), detecting the down-converted RF interferogram. The second pair is transmitted through a reference absorption cell, containing pure $CH_4$ at 100 mbar, and is dispersed by a diffraction grating. A small part of the dispersed spectrum, containing a single well-known absorption line of $CH_4$ (at 3038.498 $cm^{-1}$), is focused on a second HgCdTe photodetector (PVI-4TE, Vigo System, Poland). The outputs of the two detectors are recorded by a two channel 125 MSam/s, 16 bits, analog-to-digital convertor (NI-5762, National Instruments, US) and saved on a computer for data processing. A common 10 MHz clock is used to synchronize the dual-comb spectrometer, the data acquisition, and the modulation of the discharge current.

**Data processing and averaging.** Each of the recorded interferograms is Fourier transformed to reconstruct the spectrum. The free running carrier envelope offset frequencies and unstabilized OPO cavity length reduce the complexity of the experimental setup, but they cause a changing frequency shift in consecutive recorded spectra. Since the fluctuations of the offset frequencies and the OPO cavity length are negligible in the measurement time of a single interferogram (120 μs), i.e. the two idler combs are coherent in this time scale, a linear frequency shift is sufficient to correct for these changes in each single measurement. To address this frequency shift, we use the frequency position of the known reference absorption line and correct the frequency shift of each individual spectrum before averaging, which also yields an absolute optical frequency calibrated spectrum. The spectra are averaged after the shift correction.

**Discharge setup.** The Pyrex discharge tube is ~50 cm long, with an internal diameter of 3 mm and it is water cooled. The hollow cathodes are at the two ends of the tube and the anode is at the center. A current-stabilized high-voltage (HV) power supply (Haefely Hipotronics, US, providing up to 25 kV, 40 mA) is used for generating a DC discharge in the tube. The power supply is able to limit the current overshoots during switching/modulation of the discharge. The discharge can be switched on (glow) and off (dark) by modulating the current of HV power supply using a signal generator, whose clock is synchronized with the repetition rate different ($\Delta f_{rep}$) of the dual-comb spectrometer. The modulation signal is a square-wave whose frequency, duty cycle, and time-delay (with respect to the $\Delta f_{rep}$) can be adjusted independently. Therefore, the data acquisition and discharge process are synchronized and can also be programmed to have a time delay with respect to each other.

## Acknowledgments

This work was financially supported by Dutch Technology Foundation (NWO, 11830) and EU H2020-ICT29 (FLAIR project, 732968). The authors thank David H. Parker and Giel Berden for useful comments on the manuscript.

## Correspondence

Correspondence should be addressed to Amir Khodabakhsh (a.khodabakhsh@science.ru.nl)

## References

1    Murphy, R. E., Cook, F. H. & Sakai, H. Time-resolved Fourier Spectroscopy. *J. Opt. Soc. Am.* **65**, 600-604, (1975).

2    Uhmann, W., Becker, A., Taran, C. & Siebert, F. Time-resolved FT-IR absorption-spectroscopy using a step-scan interferometer. *Appl. Spectrosc.* **45**, 390-397, (1991).

3    Jiang, E. Y. Phase-, time-, and space-resolved step-scan FT-IR spectroscopy - Principles and applications to dynamic and heterogeneous systems. *Spectroscopy* **17**, 22-34, (2002).

4    Iwata, K. & Hamaguchi, H. O. Construction of a versatile microsecond time-resolved infrared spectrometer. *Appl. Spectrosc.* **44**, 1431-1437, (1990).

5    Yuzawa, T., Kato, C., George, M. W. & Hamaguchi, H. O. Nanosecond time-resolved infrared-spectroscopy with a dispersive scanning spectrometer. *Appl. Spectrosc.* **48**, 684-690, (1994).

6    Cunge, G., Vempaire, D., Touzeau, M. & Sadeghi, N. Broadband and time-resolved absorption spectroscopy with light emitting diodes: Application to etching plasma monitoring. *Appl. Phys. Lett.* **91**, 231503, (2007).

7    Matsugi, A., Shiina, H., Oguchi, T. & Takahashi, K. Time-resolved broadband cavity-enhanced absorption spectroscopy behind shock waves. *J. Phys. Chem. A* **120**, 2070-2077, (2016).

8    Adler, F. *et al.* Mid-infrared Fourier transform spectroscopy with a broadband frequency comb. *Opt. Express* **18**, 21861-21872, (2010).

9    Zhang, Z. W., Gardiner, T. & Reid, D. T. Mid-infrared dual-comb spectroscopy with an optical parametric oscillator. *Opt. Lett.* **38**, 3148-3150, (2013).

10   Jin, Y. W., Cristescu, S. M., Harren, F. J. M. & Mandon, J. Two-crystal mid-infrared optical parametric oscillator for absorption and dispersion dual-comb spectroscopy. *Opt. Lett.* **39**, 3270-3273, (2014).

11   Baumann, E. *et al.* Spectroscopy of the methane $\nu_3$ band with an accurate midinfrared coherent dual-comb spectrometer. *Phys. Rev. A* **84**, 9, (2011).

12   Nugent-Glandorf, L. *et al.* Mid-infrared virtually imaged phased array spectrometer for rapid and broadband trace gas detection. *Opt. Lett.* **37**, 3285-3287, (2012).

13   Galli, I. *et al.* Mid-infrared frequency comb for broadband high precision and sensitivity molecular spectroscopy. *Opt. Lett.* **39**, 5050-5053, (2014).

14   Khodabakhsh, A., Rutkowski, L., Morville, J. & Foltynowicz, A. Mid-infrared continuous-filtering Vernier spectroscopy using a doubly resonant optical parametric oscillator. *Appl. Phys. B-Lasers Opt.* **123**, 12, (2017).

15    Iwakuni, K., Bui, T. Q., Niedermeyer, J. F., Sukegawa, T. & Ye, J. Comb-resolved spectroscopy with immersion grating in long-wave infrared. *Opt. Express* **27**, 1911-1921, (2019).

16    Weichman, M. L. *et al.* Broadband molecular spectroscopy with optical frequency combs. *J. Mol. Spectrosc.* **355**, 66-78, (2019).

17    Coddington, I., Swann, W. C. & Newbury, N. R. Time-domain spectroscopy of molecular free-induction decay in the infrared. *Opt. Lett.* **35**, 1395-1397, (2010).

18    Lomsadze, B. & Cundiff, S. T. Frequency combs enable rapid and high-resolution multidimensional coherent spectroscopy. *Science* **357**, 1389-1391, (2017).

19    Kim, J., Cho, B., Yoon, T. H. & Cho, M. Dual-frequency comb transient absorption: broad dynamic range measurement of femtosecond to nanosecond relaxation processes. *J. Phys. Chem. Lett.* **9**, 1866-1871, (2018).

20    Bergevin, J. *et al.* Dual-comb spectroscopy of laser-induced plasmas. *Nat. Commun.* **9**, 6, (2018).

21    Reber, M. A. R., Chen, Y. N. & Allison, T. K. Cavity-enhanced ultrafast spectroscopy: ultrafast meets ultrasensitive. *Optica* **3**, 311-317, (2016).

22    Zhang, Y. *et al.* Time-resolved dual-comb measurement of number density and temperature in a laser-induced plasma. *Opt. Lett.* **44**, 3458-3461, (2019).

23    Fleisher, A. J. *et al.* Mid-infrared time-resolved frequency comb spectroscopy of transient free radicals. *J. Phys. Chem. Lett.* **5**, 2241-2246, (2014).

24    Bjork, B. J. *et al.* Direct frequency comb measurement of OD + CO -> DOCO kinetics. *Science* **354**, 444-448, (2016).

25    Bui, T. Q. *et al.* Direct measurements of DOCO isomers in the kinetics of OD + CO. *Sci. Adv.* **4**, 8, (2018).

26    Jin, Y. W., Cristescu, S. M., Harren, F. J. M. & Mandon, J. Femtosecond optical parametric oscillators toward real-time dual-comb spectroscopy. *Appl. Phys. B-Lasers Opt.* **119**, 65-74, (2015).

27    Abbas, M. A. *et al.* Mid-infrared dual-comb spectroscopy with absolute frequency calibration using a passive optical reference. *Opt. Express* **27**, 19282-19291, (2019).

28    Ycas, G. *et al.* High-coherence mid-infrared dual-comb spectroscopy spanning 2.6 to 5.2 mu m. *Nat. Photonics* **12**, 202-208, (2018).

29    Muraviev, A. V., Smolski, V. O., Loparo, Z. E. & Vodopyanov, K. L. Massively parallel sensing of trace molecules and their isotopologues with broadband subharmonic mid-infrared frequency combs. *Nat. Photonics* **12**, 209-214, (2018).

30    Gordon, I. E. *et al.* The HITRAN2016 molecular spectroscopic database. *J. Quant. Spectrosc. Radiat. Transf.* **203**, 3-69, (2017).

31    Draper, A. D. *et al.* Broadband dual-frequency comb spectroscopy in a rapid compression machine. *Opt. Express* **27**, 10814-10825, (2019).